# Effect of Power Losses on Self-Focusing of High Intensity Laser Beam in Gases

V. V. Semak[1] and M. N. Shneider[2]
[1] Applied Research Laboratory, the Pennsylvania State University, PA
[2] Department of Mechanical and Aerospace Engineering, Princeton University, NJ

**Abstract**
A theoretical study of power loss from periphery of an ultrashort pulse laser beam and temporally resolved defocussing produced by laser induced plasma are performed using paraxial approximation. Our analysis incorporate consideration of spatial distribution of the laser beam irradiance and the results show that substantial power losses (10%-80%) occur from the beam periphery limiting length of a filament. It was also shown that generally accepted concept of self-focusing critical power is inconsistent with consideration of self-induced refraction of spatially distributed laser beam. A new criterion for self-focusing and hypothesis for multiple filamentation are proposed.

**Introduction**

A large number of articles (currently more than 2,700) on laser beam self-focusing were published since 1962 when a concept of self-trapping of an electro-magnetic beam in media was formulated [1]. Shortly after, in 1964 in two unrelated publications [2,3] a theoretical model was proposed. This theoretical model became widely accepted as a foundation for all subsequent physical models (see reviews [4-6]). In both works [2,3] (as well as in all following works except for [7]) a self-consistent analytical solution of electromagnetic wave equation for propagation in nonlinear media was obtained under assumption that the nonlinearity induces waveguide and the beam propagation is a mode of this waveguide.

The solution for waveguide-like propagation requires specific beam intensity distribution significantly different from the Gaussian distribution that is commonly found in practical applications. This discrepancy was pointed out by Askaryan in his 1973 work [8]: "… the profile of the intensity distribution [of the beam] can be so unsuited that it will cause aberration and beam separation." Thus, self-focusing of a beam with a realistic beam profile, for example, Gaussian beam, shell be accompanied by aberration and distortion, i.e. the beam won't focus to a point as predicted by the waveguide model [3].

Another significant drawback of the existing models is that the beam refraction induced by nonlinear media response and ionization of matter is compared to the beam divergence angle due to diffraction defined as $\theta = 1.22\lambda/n_0 D$. According to strict physical definition for the laser beam divergence, it is the angle between individual ray (or the local wave vector, $k$) and the beam axis. Then, the divergence angle is dependent on the location and its value ranges from zero to infinity and, thus, it can't be set at a fixed value. A specifically set value for the beam divergence is used for engineering purposes and, depending on specific need, it is common to assign to the beam divergence angle a certain fixed value such as, for example, $1.22\lambda/n_0 D$ (D is the beam diameter), or $\lambda/\sqrt{2}\pi n_0 \sigma_0$ ($\sigma_0$ is the beam radius on 1/e intensity level), or $\lambda/\pi n_0 w_0$ ($w_0$ is the beam radius ion 1/e² intensity level). However, in physical modeling of beam propagation such engineering approach leads to conceptual misunderstandings, such as, for example, concept of "critical power" for self-focusing. The Discussion section of our article presents arguments substantiating critique of the "critical power" concept and here we would like to point out that in 1973 Askaryan expressed skepticism regarding validity of this concept in his review [8] by writing in the sub-note on the page 251: "Let's point out that name "critical power" is rather poor since the self-focusing takes place at powers below the "critical power" and manifests itself through decrease of divergence and cross section of the beam."

In the current conceptual models of high intensity beam propagation in media it is proposed [4,9] that, for some sufficient laser irradiance producing necessary electron density the plasma defocusing compensates the nonlinear self-focusing (Kerr effect) resulting in beam divergence. As the defocused laser beam propagates and the beam irradiance decreases, the plasma defocusing power decreases and the beam focusing due to Kerr effect starts dominating the beam propagation once again producing another area of beam concentration. In this new beam concentration the plasma is generated again producing beam defocussing. This balance of self-focusing compensating the divergence due to plasma and diffraction repeats multiple times maintaining the beam within certain channel, called filament. The length of the filament channel is predicted to be very large (up to several km) [4,5,9] depending on the laser pulse energy and the energy needed for gas ionization.

Unfortunately, experiments show that the length of the filament is significantly smaller than predicted. We suggest here that the discrepancy between the theoretical predictions and the experimental results is due to unaccounted previously power losses that occur from the laser beam periphery. Indeed, both Kerr self-focusing and plasma defocussing affect beam divergence less in the locations that are at larger radial distance from the beam axis. Then the beam periphery is very weakly affected by the mentioned nonlinear and plasma effects and it should diverge dominantly governed by the diffraction. The fraction of the laser beam power that is contained in this weakly affected and diverging peripheral part of the beam will be lost from participation in the filament formation and this will limit the length of the filament. It is easy to infer qualitatively that, in case of laser beams with irradiance distribution described by a monotonic, smooth function, such as for example Gaussian beam, the size of the peripheral part of the beam weakly affected by the nonlinear refraction and plasma, and thus, the amount of power loss, should be a smooth decreasing function of the laser beam maximum irradiance (or laser power) for a given beam radius. As it will be evident from the simulation results and discussion below, for the current energies of the ultrashort pulse lasers these losses are significant, limiting the filament length at the values that are substantially lower than predicted by existing theoretical models.

Performing literature search we have found only one work that considers nonlinear refraction of a laser beam as a function of radial coordinate [7]. In this work the scalar nonlinear wave equation is solved assuming slowly varying amplitude and negligible diffraction. Because the diffraction was neglected the angle of the wave vector $\mathbf{k}$ relative to the beam axis, $\theta(r,z)$ (equation (4.20) in [7]), is negative for all values of radial coordinate, $r$, (Figure 4.3 in [7]). Thus, in such approximation the beam self-focusing occurs without losses. In the same time, the computed data (Figure 4.3 in [7]) show that the parts of the beam with different radial coordinate are deflected from the direction parallel to the beam axis at different angles. From these data it is easy to see that different parts of the beam are focused into different locations along the beam axis and these "focal" locations range from some minimal value to infinity.

In relatively recent times a number of works presented results of comprehensive numerical solutions of Maxwell's equation of propagation in realistic conditions (see, for example, [10]). It is reasonable to expect that the numerical solutions should provide complete description of the beam intensity and phase evolution, in particular, showing losses from the beam periphery. However, to the best of our knowledge, there are no reports of such analysis of computed data. Thus, the only losses considered in the current physical models of self-focusing are negligibly small losses required to produce ionization.

In order to illustrate the beam wave front evolution we created a model that considers realistic conditions: non-uniform radial distribution of beam intensity, focused beam, nonlinear refractive index component that is assumed to be instantaneous, component of refractive index that is dependent on local density of free electrons that is evolving in time. For the sake of simplification and without jeopardizing predictive accuracy and illustrative strength we devised an integrated physical and geometrical optics approach. This approach combines known analytical solution of Maxwell's equation for a Gaussian beam propagation in linear media [11] with the solution of the ray-tracing (eikonal) equation under assumption of paraxial propagation [12]. This approach provides elegant way for numerical computation of the ray trajectories (avoiding singularities) as the focused laser beam propagates in a nonlinear and ionized media through its caustic (the area near the focal plane with the length of several Rayleigh lengths).

**Theoretical model**

The expression for the refractive index, $n$, is as follows

$$n = n_0 + n_2 I - \omega_p^2 / 2\omega_0^2, \quad \omega_p^2 = \frac{e^2 N_e}{\varepsilon_0 m_e}, \tag{1}$$

where $n_0$ is the linear index of refraction in air; $n_2$ and $I$ being the coefficient of the Kerr nonlinear index of refraction and the local laser beam irradiance, respectively; $\omega_p$ is the plasma frequency; $N_e$ is the density of electrons produced due to ionization, $m_e$ is the mass of electron and $\omega_0$ is the frequency of laser.

The equation for the angle of individual ray (angle between the wave vector and the beam axis), $\varphi$, after propagating length $ds$ along its trajectory in a media with nonuniform refractive index, $n$, can be derived from the eikonal equation [12]

$$\frac{d}{ds}\left(\frac{d\vec{r}}{ds} n\right) = \nabla n, \tag{2}$$

where $\vec{r}$ is the radial vector, $\varphi = \arcsin\left(\frac{dr}{ds}\right)$, and $dr$ is change of the radial vector along the coordinate normal to coordinate z that coincides with the beam axis. Here by convention the positive value of the angle of an individual ray corresponds to divergence (wave vector is directed away from the beam axis) and the negative value corresponds to convergence (wave vector directed toward the beam axis). In paraxial approximation that assumes small $\varphi$ and, therefore, $ds = dz/\cos\varphi \approx dz$, it follows from the equation (2)

$$d\varphi^{NLP} \approx \frac{\partial n/\partial r}{n} dz, \tag{3}$$

where $d\varphi^{NLP}$ is the change of the angle of individual ray due to Kerr self-focusing and plasma defocussing, $z$ is the coordinate along the beam axis $r$ is the radial coordinate normal to the beam axis, and the refractive index, $n$, is given by the equation (1).

For the diffraction dominated beam propagation and under assumption of paraxial propagation the angle of an individual ray, i.e. angle of the wave vector relative to the beam axis, $\varphi^D(r,z)$, can be approximated [11] as a ratio of radial position $r$ and the local wave front curvature radius, $R(z)$:

$$\varphi^D(r,z) \approx \frac{r}{R(z)}. \tag{4}$$

In this work we consider a single mode (Gaussian) beam then the wave front curvature radius is given by the equation [11]

$$R(z) = z\left(1 + \left(\frac{\pi w_0^2}{\lambda z}\right)^2\right), \tag{5}$$

where, $w_0$ is the beam radius on the $1/e^2$ level of the maximum beam irradiance in the beam waist and $\lambda$ is the laser wavelength. According to the convention, $z$ axis coordinate is set to zero at the beam waist and for a beam propagating from left to right, the $z$ values are negative to the left and positive to the right from the waist location. Thus, the diffraction dominated beam is converging for coordinates $z<0$ and the radius of the wave front curvature is negative and it is diverging for $z>0$ and the radius of the wave front curvature is positive.

Assuming that the *r*-coordinate of an individual ray is controlled by the diffraction only and, thus, the contribution to change of *r*-coordinate by the refraction due to nonlinear effect is neglected, the propagation of an individual ray from location with coordinate $z=z_0$ to the location with coordinate $z=z_s$ can be described in terms of angle of individual rays (the angle of wave vector), $\varphi^{NLP}(r,z_s)$, represented by a function of variables $r$ and $z_s$ obtained as a result of integration of the equation (3) for a given initial conditions, $\varphi_0(r, z = z_0)$:

$$\varphi(r,z_s) = \varphi^D(r,z_s) + \varphi^{NLP}(r,z_s) = \varphi^D(r,z_s) + \int_{z_0}^{z_s} d\varphi^{NLP} = \varphi^D(r,z_s) + \int_{z_0}^{z_s} \frac{\partial n/\partial r}{n} dz, \tag{6}$$

Let's assume that the laser beam is focused and the initial condition is chosen in the location $z=z_0$ between the lens and the beam waist where the laser beam irradiance is too small to produce nonlinear effect, i.e. only diffraction divergence is responsible for the beam evolution. Then under these assumptions, it follows from the equations (4,5) following expression for the initial condition:

$$\varphi_0(r, z = z_0) = \varphi^D(r, z = z_0) \approx \frac{r}{z_0 \left(1 + \left(\frac{\pi w_0^2}{\lambda z_0}\right)^2\right)}. \qquad (7)$$

In order to solve the propagation equation (6) under conditions when nonlinear effects become non-negligible, we need to use the expression for the refractive index given by the equation (1) and, therefore, a functions $I(r,z,t)$ describing the beam irradiance and $N_e(t)$ describing number density of electrons are needed. The beam irradiance distribution can be computed from Maxwell's equations that must be solved in self-consistent way and this is a very difficult problem. For the sake of simplification that does not diminish generality of the obtained results and still is capable of providing good accuracy of computations we assume that, the beam has undisturbed Gaussian distribution in the volume where the nonlinear refraction occurs, i.e. in the near filed there is negligible distortion of the beam profile. Note that this assumption is equivalent to the assumption used to formulate the equation (6). Of course, in the far field the small changes of the angles of the wave front vector result in substantial changes of beam profile.

Thus, the beam irradiance within the computational range of z-coordinate is

$$I(r, z, t) = \frac{2P(t)}{\pi w^2(z)} \exp\left(-\frac{2r^2}{w^2(z)}\right), \qquad (8)$$

where $P(t)$ is time dependent laser power and $w(z)$ is the laser beam radius at the intensity level of $1/e^2$ at the location along the beam axis with the coordinate z that is given by the equation

$$w^2(z) = w_0^2 \left(1 + \left(\frac{\lambda z}{\pi w_0^2}\right)^2\right). \qquad (9)$$

It is easy to see that typical ultrashort pulse lasers wavelengths and pulse energies and interaction with air the ultrashort laser pulses produce multiphoton ionization ($\gamma \gg 1$) with the electron number density given by the equation

$$N_e(r, z, t) = \sigma_k (N_0 - N_e) \int_{-\infty}^{t} [I(r, z, \xi)]^k d\xi, \qquad (10)$$

where $N_0$ is the number density of neutrals, $\sigma_k$ is the ionization rate due to absorption of $k$ photons such that $k = \text{int}\left[I_i/\hbar\omega\right] + 1$, where $I_i$ is the potential of ionization of gas and $\omega$ is the laser angular frequency.

**Computational procedure and results**

The propagation equation (6) was solved numerically. Every time prior to performing computations for a selected set of input data we verified that, 1) the coordinate $z=z_0$ where integration of equation (6) starts is selected outside of the volume where beam intensity is sufficient to produce noticeable non-linear refraction and 2) the maximum radial displacement of the individual rays after propagating from the location $z=z_0$ to any location within the volume where nonlinear refraction is substantial does not exceed 5% of the beam radius $w(z)$ in this location.

The verification of condition 1) was performed by integrating the propagation equation (6) from a different initial coordinate (negative value) $z_0$ to a fixed final coordinate $z_s$ (positive value) and determining the smallest value of $z_0$ (most distant in the negative direction from the beam waist) when the output function $\varphi(r, z_s)$ changing becomes negligibly small. This value of $z_0$ was used for the initial coordinate and the final coordinate was set at $z_s = -z_0$. It was observed that for the all conditions used the integration range $[z_0; z_s]$ was within $[-2z_R; 2z_R]$, where $z_R$ is the Rayleigh length: $z_R = \dfrac{\pi w_0^2}{\lambda z_0}$.

The verification of condition 2) was performed by integrating equation for the radial displacement of individual ray:

$$\Delta r(r) = \int_{z_0}^{z_s}\left(\int_{z_0}^{z_s} d\varphi^{NLP}\right) dz = \int_{z_0}^{z_s}\left(\int_{z_0}^{z} \dfrac{\partial n/\partial r}{n} dz\right) dz . \qquad (11)$$

Then, the maximum value of normalized radial displacement of a ray, $\max\left\{\Delta r(r)/w(z_s)\right\}$, was compared with the corresponding dimensionless beam radius $r/w(z_s)$. Only those computations in which the maximum normalized ray displacement was less than 5% of the dimensionless beam radius were accepted.

The computations were performed for a laser wavelength $\lambda = 800$ nm and a Gaussian laser beam with the collimated beam radius on $1/e^2$ level, $w_1 = 3$mm, focused by a lenses with the focal lengths $F$ in the range 0.0375 m - 1.875 m, that corresponds to the beam waist radii, $w_0$, in the range 10 μm – 500 μm. Further, for the calculation of index of refraction of air we used the coefficient of the Kerr nonlinear index of refraction $n_2 = 5 \cdot 10^{-23}$ m$^2$/W [13] and for the multi-photon ionization calculations, assuming 8 quanta absorption, we used $\sigma_8 \approx 1.5 \cdot 10^{-130}$ m$^{16}$/s/W$^8$ extrapolated from the computed data [14].

We assumed that the main ion produced in air is $O_2^+$. The Keldysh parameter [15] $\gamma = \frac{\omega\sqrt{2mI_i}}{eE}$, for oxygen molecules $I_i = 12.2$ eV. For highest considered laser power, P=1.5 GW, and $w_0 = 5 \cdot 10^{-5}$ m, the laser intensity, $I \approx 2P/\pi w_0^2 = 3.82 \cdot 10^{17}$ W/m$^2$ and the corresponding amplitude of electric field, $E \approx 1.7 \cdot 10^{10}$ V/m. Therefore, the Keldysh parameter $\gamma \approx 1.62$ and the limit of multiphoton photoeffect is valid.

The results of simulation of a laser beam propagation through air, assuming a rectangular shape laser pulse with duration τ = 100 fs, and a Gaussian laser beam with the collimated beam radius on $1/e^2$ level, $w_l$ = 3mm, focused by a lenses with the focal lengths F= 0.589 m that produces beam waist $w_0$ = 50 μm, are sown in the Figure 1. The angle of the wave vector, $\varphi_{z_s}$ computed for the beam of various total power, $P_0$, after propagating from $z_0$ = -2$z_R$ to $z_s$ = 2$z_R$ ($z_R$=19.62 mm) is shown as a function of dimensionless radius $r/w(z_s)$ at the end of the 100 fs laser pulse assuming a rectangular pulse shape. Simultaneously, for the sake of reference, the ratio of the laser power passing through a circle with radius $r$ to the total laser power, $P_r/P_0$, (right axis) is shown as a function of dimensionless beam radius $r/w(z_s)$.

For the low laser power the angle $\varphi_{z_s}$ depends linearly on the dimensionless beam radius, i.e. the wave front is spherical as expected when only the diffraction governs the beam propagation. The increase of laser power results in higher contribution of the nonlinear part of the refractive index (second term in the right hand side of the equation (1)) producing focusing of the central part of the beam. For lower powers this focusing contribution is insufficient to overcome the beam divergence due to diffraction simply reducing the angle of divergence of the beam in relatively small central part of the beam. As the laser power is increased the compensation of the divergence increases (more substantially near the beam axis) and the affected area of the near axial part of the beam increases. At certain laser power, that we will denote as "self-focusing power" or $P_{SF}$, the self-focusing compensates the diffraction divergence of the beam in some small area near the beam axis – the area where $\varphi_{z_s}$ = 0. For the laser powers that exceed $P_{SF}$ further increase of the laser power produces negative angles in the near axis area of the beam, i.e. the beam is converging in this area. The amplitude of the angle of convergence (the lowest negative value of the angle $\varphi_{z_s}$) and the area of the beam where the wave vector angle is negative are increasing with the increase of laser power and, thus the larger and larger part of laser beam is focused with larger and larger convergence angle.

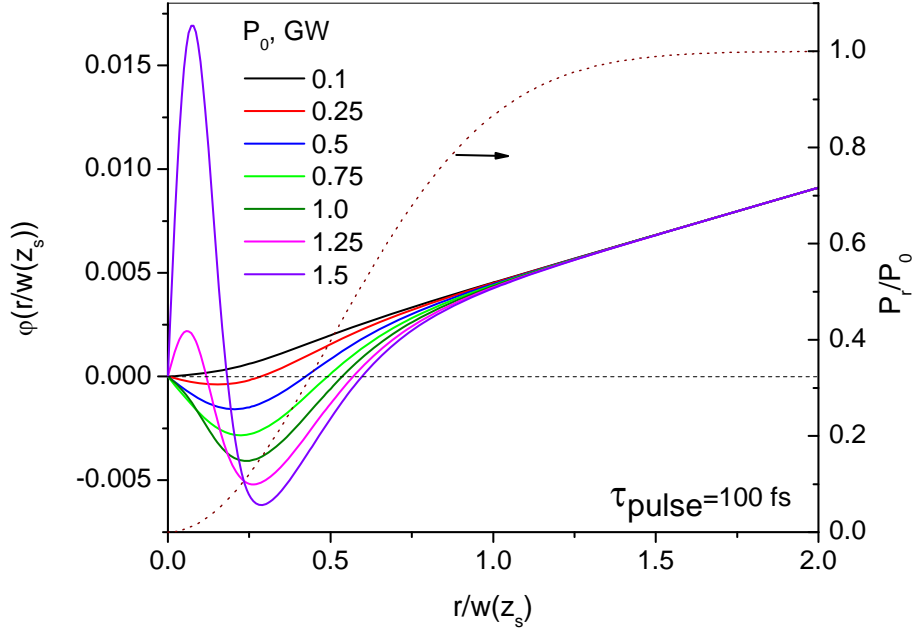

Fig. 1. The angle of individual rays of a collimated Gaussian beam with 3mm radius at $1/e^2$ intensity level focused by a lens with focal length $F = 0.589$ m computed as a function of relative beam radius for different maximum power of a rectangular shape 100 fs laser pulse propagating in air. The computed angle dependence is shown for time at the end of the laser pulse and at the location $z_s = 2z_R = 19.62$ mm. The computations of the beam propagation were performed from the initial location $z_0 = -2z_R = -19.62$ mm.

As is evident from the Figure 1, for the laser powers that exceed $P_{SF}$ the part of the laser beam with relative radius in the range from 0 to the value where the function $\varphi_{z_s}\left(r/w(z_s)\right)$ has minimum, the convergence angle $\varphi_{z_s}$ decreases almost linearly with increase of relative radius of the beam. The behavior of this part of the beam is as it is focused by a lens into a focal spot located at approximately $z = F_{SF} = \dfrac{r_{min}}{\varphi_{z_s}\left(r_{min}/w(z_s)\right)}$, where $r_{min}$ is the radius where the angle $\varphi_{z_s}$ is at its minimum. Thus, the value $F_{SF}$ is positive and can be considered as approximate focal length associated with the Kerr effect.

It is also easy to see that the part of the beam with relative radius in the range from the value where the function $\varphi_{z_s}\left(r/w(z_s)\right)$ has minimum to the value where $\varphi_{z_s} = 0$, is focused into a volume that is spread along the axis of the beam from $z_F$ to infinity in such a way that the part of the beam with radial coordinate closer to the beam axis is focused into a spot with the smaller $z$ coordinate.

The results of our simulations of the focused laser beam propagation in air demonstrate that only some near axis fraction of the beam is affected by self-induced refraction. The peripheral part of the beam that carries substantial amount of power remains unaffected and diverges governed by the diffraction. This is demonstrated by the computational example shown in the Figure 1, that shows that the beam periphery where relative radius exceeds approximately unity, $r/w(z_s) > 1$, diverges unaffected by the nonlinear processes containing approximately 14% of total power. This peripheral part of the beam has angles exceeding those in the near-axis part of the beam and it doesn't cross with the near-axis part. Thus, unaffected peripheral part carries away irretrievably approximately 14% of the total laser power representing losses.

It is reasonable to propose more stringent loss criterion declaring that, the part of the laser beam where $\varphi_{z_s} > 0$ contains power lost as far as self-induced refraction is concerned. At this point we disregard contribution to losses from the near-axis zone where substantial defocusing by plasma occurs for higher laser powers (Figure 1) as it will be discussed below. The results shown in the Figure 1 demonstrate that, assuming this more stringent criterion, the fraction of the lost beam power is approximately 45% for laser power 1.5 GW and it is increasing with decrease of laser power: for example, more than 80% of total power does not participate in self-induced refraction (i.e. is lost) for 0.25 GW laser power.

The results presented in the Figure 1 also show that, if the laser power exceeds some value $P_{ION}$ that is in the range 1 GW – 1.25 GW in particular case, a function $\varphi_{z_s}\left(r/w(z_s)\right)$ becomes positive in the very close near-axis area with a large amplitude maximum in addition to the described above negative angle area with minimum produced due to nonlinear Kerr self-focusing. Obviously, for the particular laser parameters, the increase of laser power brings into play the third term in the equation (1) that describes the contribution to the refractive index provided by the ionization. Because this contribution is negative and absolute value of this term is a decreasing function of the relative beam radius, plasma is responsible of defocussing of the laser beam. Unlike Kerr effect that can be considered as almost instantaneous, the defocussing due to ionization increases during the laser pulse as the electron number density $N_e$ grows.

Predicted strong defocusing due to refraction in plasma possibly contributes to the previously discussed losses. However, without comprehensive computations it is impossible to assess the amount of the loss induced by the plasma refraction. Plasma is produced in the area of maximum beam intensity and it refracts this high intensity portion of the beam at comparatively large angle. This can create a ring-shaped off-axis zone of high beam intensity where secondary self-induced refraction can occur. We propose here a hypothesis, that this is the mechanism that produces multiple filaments: when laser power exceeds certain level, the multiphoton ionization results in sufficiently large electron number density that produces strong refraction of the near-axis high intensity that, in turn, creates secondary ring–shaped zone of high intensity where the nonlinear process of self-focusing and multiple ionization produces secondary filaments. Some instability is responsible for "breaking" of this ring structure into multiple stochastically distributed and shaped secondary filaments.

The temporal dynamics of the radial distribution of rays in the laser beam affected by the Kerr self-focusing and plasma defocusing is shown in the Figure 2 for the interaction conditions that are the same as in Figure 1. The Kerr effect has negligibly small characteristic rise time and, therefore, in our consideration it is treated as instantaneous. Since we assumed for the sake of simplicity that the laser pulse has a rectangular temporal profile the self-focusing effect starts immediately at the beginning of laser pulse and doesn't change during the laser pulse. On the contrary ionization is a process that has characteristic time comparable to the laser pulse duration. The ionized volume starts noticeably contributing to the beam refraction at the later stages of the laser pulse at time that is the range between 10% and 20 % of the laser pulse duration. The angle $\varphi_{z_s}\left(r/w(z_s)\right)$ increases almost linearly in the small near axis zone where $r/w(z_s) < R_{max}$, reaching positive maximum value that is growing in time. Thus this part of the beam is defocussed by an ionized gas lens with negative focal length, $F_{ION}(t) = -r_{max}(t)/\varphi_{z_s}^{max}(t)$, that has increasing in time absolute value.

For the relative beam radius larger than $R_{max}$ the angle $\varphi_{z_s}\left(r/w(z_s)\right)$ decreases almost linearly until it, first, becomes zero for the relative radius equal to $R_{zero}^1$, and then, reaches its minimum when the relative radius is $R_{min}$. The part of the beam in this range of relative radius where the angle is linearly decreasing and is positive corresponds to a defocussing by a "lens" with variable (as a function of relative radius) negative focusing length $F_{def}^1 = -r/\varphi_{z_s}\left(r/w(z_s)\right)$ that changes from approximately $F_{ION}$ to zero as relative radius changes from $R_{max}$ to $R_{zero}^1$. The part of the beam for the relative radius in the range from $R_{zero}^1$ to $R_{min}$, where the angle of wave front is almost linearly decreasing as a function of relative radius and is negative corresponds to the focusing with an O-ring shaped lens with focal length $F_{SF+ION}(t) = r_{min}(t)/\varphi_{z_s}^{min}(t)$. It is easy to see that this lens will focus the laser beam into a focal spot that has O-ring shaped cross section.

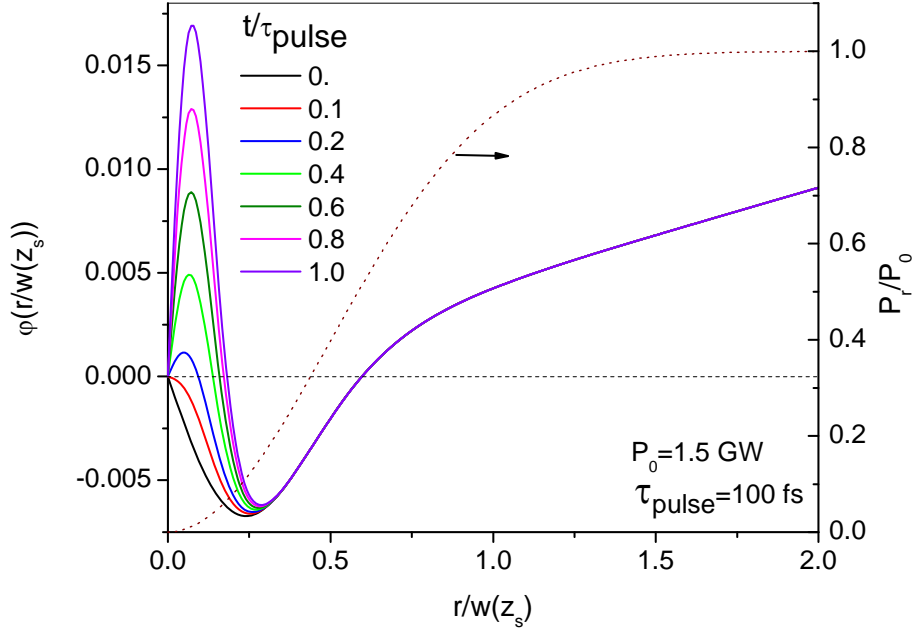

Fig. 2. The temporal evolution of the angle of individual rays of a collimated Gaussian beam with 3mm radius at $1/e^2$ intensity level focused by a lens with focal length F = 0.589 m computed as a function of relative beam radius for laser power $P_0$=1.5GW and rectangular shape 100 fs laser pulse propagating in air. The computed angle dependence is shown at the location $z_s$= 2 $z_R$ = 19.62 mm. The computations of the beam propagation were performed from the initial location $z_0$ = -2 $z_R$ = -19.62 mm.

As discussed above, for the relative radius values larger than $R_{min}$ the angle of wave front vector increases again until the angle second time becomes zero at $r/w(z_s) = R^2_{zero}$, and, then becomes positive growing value for $r/w(z_s) > R^2_{zero}$. In this range of relative radius values between $R_{min}$ and $R^2_{zero}$ the beam is focused with a focusing lens that has a variable (as a function of relative radius) positive focusing length $F^1_{foc} = \dfrac{r}{\varphi_{z_s}(r/w(z_s))}$ that changes from approximately $F_{SF+ION}$ to zero. For the relative radius values exceeding $R^2_{zero}$, the beam diverges with the divergence is partially compensated by self-focusing. This self-focusing compensation decreases with the radius increase until for the relative radii larger than approximately unity the beam divergence is diffraction dominated, i.e. where the wave front angle is a positive linearly growing function of relative radius.

**Discussion**

The computational results shown in the Figure 1 are schematically presented in the Figure 3a,b. For the laser powers that produce negligible gas ionization (laser power is smaller than $P_{ION}$) and are sufficiently high in order to produce noticeable nonlinear optical Kerr effect

(laser power exceeds $P_{SF}$) the angle of the wave vector is nearly linearly decreasing function of the radial position of the ray in the near axis region of the beam (Figure 1). The rays in this part of the beam, shown with green color in the Figure 3.a, significantly deviate from the diffraction prescribed path, shown as dotted line, and because of almost linear dependence of the refraction angle on the radial coordinate, the rays converge in a way that is similar to focusing by a lens with the focal length of approximately $F_{SF}$.

    The part of the beam with larger radial coordinate, shown with orange color in the Figure 3.a, is also strongly affected by the nonlinear Kerr effect, however, to comparatively lesser degree. Although, the angle of the rays in this part of the beam is negative, i.e. the rays are converging, the angle increases as function of the beam radius. Thus, in this part of the beam the rays located at larger radial distances are "focused" into the spots positioned further along the beam axis. At certain radial location the Kerr self-focusing exactly equal to the diffraction divergence and the rays located at such radial distance are parallel to the beam axis (shown with light blue color in the Figure 3.a). At this radial position the computed angle $\varphi_{z_s}$ equals to zero (Figure 1). For the rays with larger radial distance the refraction angle is positive, i.e. the rays are diverging; however the divergence angle is still lesser than the diffraction divergence because of noticeable Kerr focusing. The deviation from the diffraction prescribed path decreases with the increase of the radial position of a ray until the Kerr self-focusing becomes negligible (as for the radial distance of approximately $w_0$ shown in Figure 3.a).

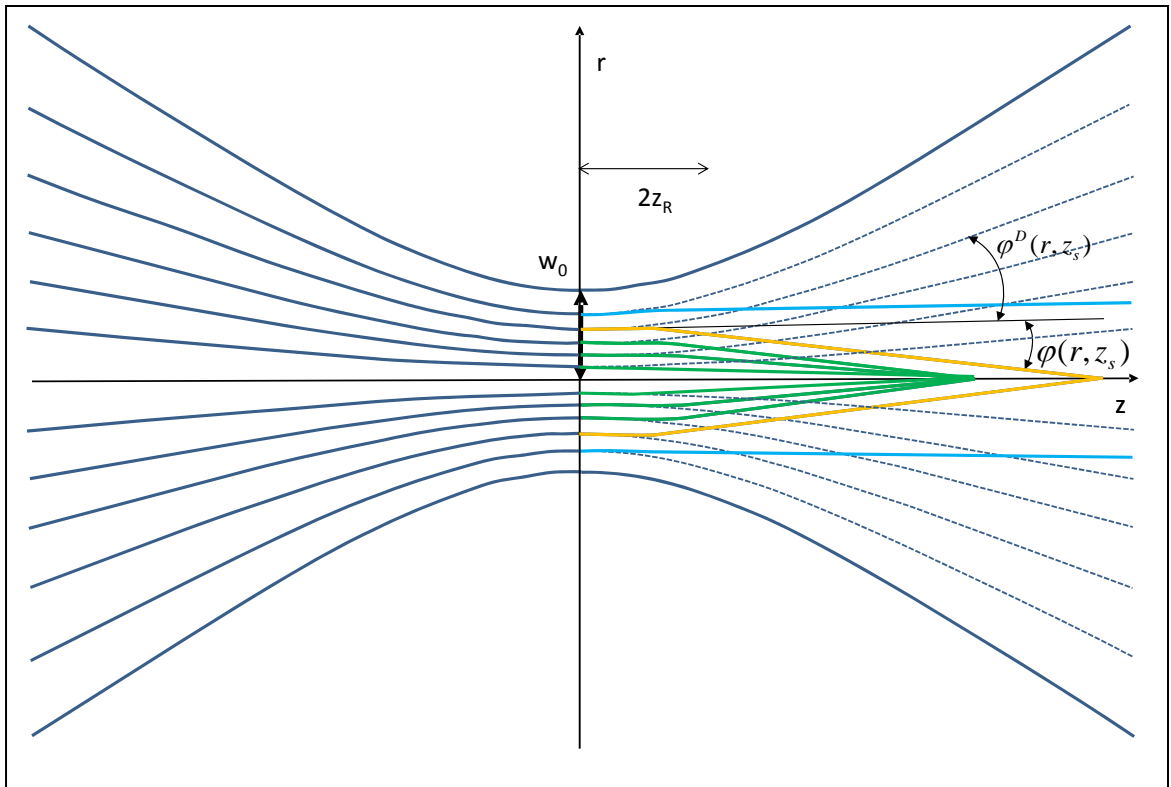

Fig.3.a. Schematic interpretation of the computational results shown in the Figure 1 for the laser beam power below $P_{ION}$.

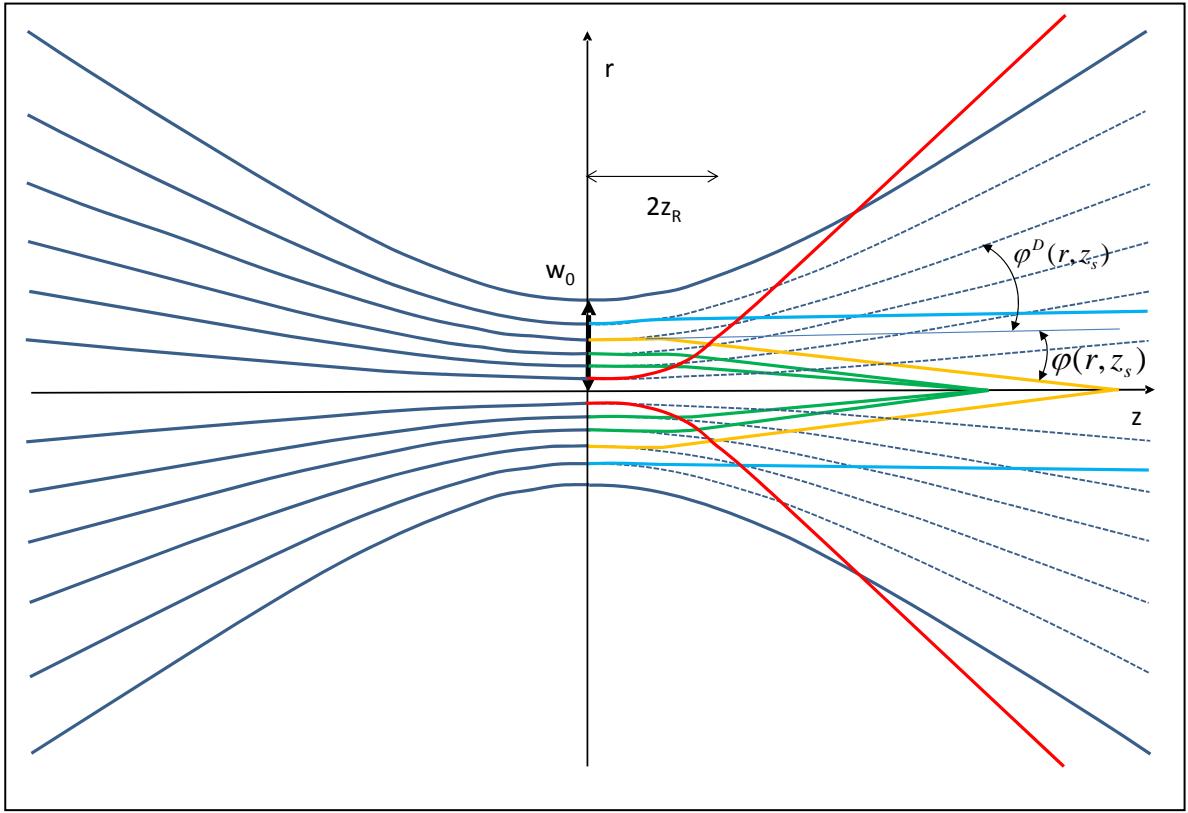

Fig.3.b. Schematic interpretation of the computational results shown in the Figure 1 for the laser beam power equal or exceeding $P_{ION}$.

The rate of ionization of gas due to multi-photon absorption or tunneling effect is strongly dependent on the laser beam maximum intensity and the substantial ionization occurs when laser power exceeds some value $P_{ION}$ for a given beam size. Thus, although the laser induced ionization is not a threshold process, it appears as a threshold like process. Because of the strong dependence of ionization rate on laser intensity, the gas ionization that noticeably affects the beam is formed first in a small near-axis region. Also, as discussed above the ionization rate increases during the laser pulse until it reaches maximum at the end of the pulse (Figure 2). The near-axis ionized volume produces defocussing shown with red rays in the Figure 3.b. According to the computed results (Figure 1), the defocusing angle produced by the ionized volume increases as function of laser irradiance. For example, for the laser power of 1.5GW and the beam focusing conditions as in the Figure 1, the divergence angle produced by the refraction in the ionized volume exceeds the angle of divergence due to diffraction within the area of the beam with radial distance of approximately $3\,w(z_s)$. Also, for the 1.5GW power of the laser beam, the radial dimension of the ionized volume refracting the near-axis rays is approximately $0.2\,w(z_s)$ containing approximately 10% of total laser power (Figure 1). This near–axis part of the beam diverged by the ionized volume carrying significant amount of laser power intersects with the peripheral parts of the beam focused due to the Kerr effect as shown in the Figure 3.b. Our hypothesis is that when a certain laser power is exceeded the amount of power in the ionization diverged part of the beam becomes sufficiently large and intersection with the focused peripheral part of the beam produces high irradiance spots that produces secondary filaments (mutifilamentation).

Thus, several important conclusions follow from the analysis of our results. First, our theoretical results demonstrate drawback of the concept of critical power introduced in early 1960s-1970s [3,6] that does not concern with the beam intensity profile. We propose that it is logical to define self-focusing as when, due to the Kerr effect, at least part of the laser beam wave front after propagating in the nonlinear media becomes collimated or converging instead of diverging due to diffraction, i.e. the wave vector angle $\varphi(r, z_s) \leq 0$ for at least some part of the beam. The diffractive divergence is compensated stronger for higher laser beam power and, then, we propose to define the threshold power of the laser beam self-focusing, $P_{SF}$, as power when the wave front angle in some part of the beam becomes zero after propagating some distance through the media. For example, the application of such definition can be illustrated by the Figure 1: the self-focusing power, $P_{SF}$, when the function $\varphi(r, z_s)$ "touches" the horizontal axis. Such defined self-focusing laser beam power was computed as a function of the radius of laser beam in the waist, $w_0$, assuming no defocusing due to ionization takes place (see Figure 4). For the beam waist radius values lesser than 50 μm the self-focusing power $P_{SF}$ is a rapidly decreasing function and for the beam radii larger than approximately 50 μm - 100 μm it is very slowly increasing (practically constant) function. Here we shell stress that the results shown in the Figure 4 represent only the self-focusing contribution and for small radius values (lesser than 50 μm) the beam intensity corresponding to the laser power larger than $P_{SF}$ is sufficient to produce air ionization that will defocus laser beam, as seen in the Figure 2.

The values of the defined above self-focusing power, $P_{SF}$, are much lower than the critical power of self-focusing, $P_{cr}$, introduced in [3,6]. In case of air and for the laser wavelength of 800 nm the critical power $P_{cr} = 3.72 \lambda_0^2 / 8\pi n_0 n_2 \approx 1.7 \text{GW}$ and its value is independent of the beam radius. The difference between the definitions stems from the dissimilarity of the theoretical concepts. The expression for critical power is deduced within concept that assumes total internal reflection on the boundary between the beam peripheral area where the beam intensity is low and with relatively lower refractive index, $n_0$, and the core of the beam where beam intensity is high and the refractive index is relatively larger (see the equation (1)). This concept is essentially describes self-focusing as a waveguide channeling within the waveguide that is produced by a uniform beam with a sharp edge. Our concept describes self-focusing as a refraction of the beam on self-induced gradient of refractive index under conditions that all characteristic lengths (length of refractive index change and beam radius) are comparable. We believe it is more accurate representation of self-induced refraction by Gaussian and Gaussian like multimode beams.

It is reasonable to suggest that because of smaller divergence the central part of a realistic beam would separate from the peripheral part (that will be irretrievable lost) and behave similarly to "hat top" beam considered in [3] producing channeling. However, the diffraction blurs the sharp edges on the characteristic length $L \sim D^2 / \lambda$, where D is the characteristic diameter. This length for the beam with 1 mm diameter of the central part is approximately 1 m. After propagating this distance the beam intensity distribution will become "bell" shaped that is more accurately described by the above described model.

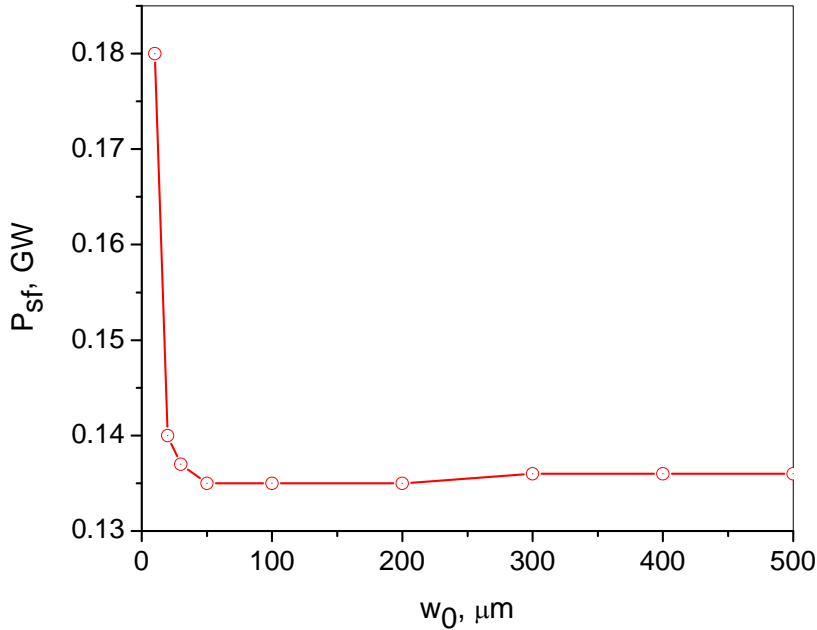

Fig. 4. The laser power corresponding to the initiation of self-focusing computed as function of the Gaussian beam waist radius at $1/e^2$ intensity level for rectangular shape 100 fs laser pulse propagating in air.

Second, our theoretical and numerical results demonstrate the role and amount of power losses from the periphery of the laser beam. According to our computations shown in the Figure 1, the losses from the peripheral part of the beam are substantial and for the realistic beam power values in the range 0.25 GW – 1.5 GW these losses (depending on the rigor of criterion discussed above) are in the range 10% - 80%. It is previously discussed [4,5,16] that the high intensity laser beam can experience repeated self-focusing. Under the concept of channeling of a beam with power exceeding $P_{cr}$, the losses from the beam periphery are neglected leading to expectation of unlimited filament length. In contrast, from our model it follows that the number of the self-focusing events and, thus, the filament length is limited due to always present significant losses.

The non-ionization divergence related losses can be reduced by placing the laser beam into the waveguide. Such waveguide propagation reduces the divergence of the laser beams in the interference region, formed by the intersection of two almost parallel laser pulses [17]. As a result of the interference regions, enhanced ionization is produced. These areas of increased ionization represent the waveguides "walls" for laser beams that could suppress their divergence, resulting finally in dramatically increasing of the attainable degree of ionization.

Third, it was shown that the increase of the gas ionization results in initiation of defocusing of the near-axis part of the beam that is smaller than the part of the beam affected by nonlinear self-focusing. Thus, there is no balance between the self-focusing and "plasma" defocusing that is commonly assumed to be a mechanism for long filament formation [5].

The ionization related defocusing from the near-axis area of the beam adds to the power losses. However, this defocusing can create areas of increased intensity that have circular shape and located where the defocussed rays intersect with the rays of the self-focused part of the beam and, possibly with the rays of the peripheral part of the beam that are diverging due to the domination of the diffraction. These high intensity areas can create conditions for self-focusing when required intensity and intensity gradient are reached, producing secondary filament of circular shape. We hypothesize that the fluctuation of density, non-symmetric beam shape and other fluctuations result in separation of this secondary circular filament into a number of multiple filaments. We also propose that a necessary condition for multiple filamentation is sufficient gas ionization, i.e. generation of sufficiently high electron number density that produces significant defocusing of near-axis part of the laser beam. Prediction of the number, location, and shapes of the secondary filaments requires further theoretical and numerical studies; however, presented discussion demonstrates that the current criterion [5,18] that defines the number of filaments as ratio of laser power to critical power, $P_{cr}$, is flawed.

**Conclusions**

In this work we performed a theoretical study of self-focusing and plasma defocussing of laser beams in paraxial approximation. The computed results demonstrate that laser beam divergence is affected by Kerr focusing and plasma defocusing differently in different radial locations of the laser beam and, therefore, adequate analysis must consider spatial profile of the laser beam. Hence, the criterion of critical power introduced in early 1960s [3,6] that is not concern with the beam intensity profile is flawed. Instead, we propose a criterion of self-focusing power that is: when due to the Kerr effect, at least part of the laser beam wave front after propagating in the nonlinear media becomes collimated or converging instead of diverging due to diffraction, i.e. the wave vector angle $\varphi(r, z_s) \leq 0$ for at least some part of the beam. Application of this new criterion shows that the Kerr self-focusing noticeably affects laser beam at powers that are by order of magnitude lower than previously proposed critical power.

Also, consideration of the nonlinear optical effects for realistic beam irradiance profiles leads to awareness of significant power losses as far as non-linear self-induced refraction is concerned. This previously disregarded power loss from the peripheral part of the laser beam weakly affected by non-linear optical effects can be significant ranging from 10% for high power to 80% for low powers that just exceed the above defined self-focusing power. Thus, filament length is substantially limited as compared to predictions of the previous models.

Finally, computations showed that at certain power levels (that exceed self-focusing power) the near-axis part of the laser beam carrying high irradiance is strongly defocussed by the plasma that forms during the laser pulse. We hypothesize that this defocussing from the high irradiance near-axis part of the beam combined with the self-focusing from the off-axis part of the beam produces ring-shaped areas of high laser irradiance generating secondary filament. Conceivable instabilities, that will be studied in future, are responsible for the breaking of this ring-shaped secondary filament into multiple filaments.